\documentclass[12pt]{article}
\PassOptionsToPackage{hyphens}{url}
\usepackage[pagebackref=true,colorlinks]{hyperref}
\usepackage{amsmath,amssymb,amsthm,mathtools,amsrefs,mathrsfs}
\usepackage{pifont}
\usepackage{mathpazo} 
\usepackage{eulervm}
\usepackage{csquotes}
\usepackage{lscape}
\usepackage{etex} 
\usepackage[table]{xcolor}
\usepackage{tikz}
\usetikzlibrary{arrows.meta}
\usetikzlibrary{decorations.pathmorphing,shapes}
\usetikzlibrary{calc}
\usepackage{comment}
\usepackage{adjustbox}
\usepackage{scalerel,stackengine}
\usepackage{stmaryrd}
\usepackage{fancyhdr}
\usepackage{soul}
\usepackage{dsfont}
\usepackage[normalem]{ulem}
\usepackage{longtable}
\usepackage[bottom]{footmisc}
\usepackage{graphicx}
\usepackage{caption}
\usepackage{subcaption}
\usepackage{pictexwd, dcpic}
\usepackage{float}
\usepackage{enumerate}
\usepackage[all]{xy} 
\usetikzlibrary{circuits.ee.IEC}
\hypersetup{
	colorlinks=true,
    linktoc=all,
    linkcolor=blue,  
    citecolor=blue,
    }

\usepackage{tocbasic}
\DeclareTOCStyleEntry[
  beforeskip=.2em plus 1pt,
  pagenumberformat=\textbf
]{tocline}{section}

\makeatletter
\newcommand*{\shifttext}[2]{%
  \settowidth{\@tempdima}{#2}%
  \makebox[\@tempdima]{\hspace*{#1}#2}%
}
\makeatother
\makeatletter
\renewcommand*\env@matrix[1][\arraystretch]{%
  \edef\arraystretch{#1}%
  \hskip -\arraycolsep
  \let\@ifnextchar\new@ifnextchar
  \array{*\c@MaxMatrixCols c}}
\makeatother

\stackMath
\newcommand\reallywidehat[1]{%
\savestack{\tmpbox}{\stretchto{%
  \scaleto{%
    \scalerel*[\widthof{\ensuremath{#1}}]{\kern.1pt\mathchar"0362\kern.1pt}%
    {\rule{0ex}{\textheight}}
  }{\textheight}%
}{2.4ex}}%
\stackon[-6.9pt]{#1}{\tmpbox}%
}
\makeatletter
\pgfmathdeclarefunction{erf}{1}{%
  \begingroup
    \pgfmathparse{#1 > 0 ? 1 : -1}%
    \edef\sign{\pgfmathresult}%
    \pgfmathparse{abs(#1)}%
    \edef\x{\pgfmathresult}%
    \pgfmathparse{1/(1+0.3275911*\x)}%
    \edef\t{\pgfmathresult}%
    \pgfmathparse{%
      1 - (((((1.061405429*\t -1.453152027)*\t) + 1.421413741)*\t 
      -0.284496736)*\t + 0.254829592)*\t*exp(-(\x*\x))}%
    \edef\y{\pgfmathresult}%
    \pgfmathparse{(\sign)*\y}%
    \pgfmath@smuggleone\pgfmathresult%
  \endgroup
}
\makeatother

\theoremstyle{theorem}
\newtheorem{theorem}[equation]{Theorem}
\newtheorem{lemma}[equation]{Lemma}
\newtheorem{proposition}[equation]{Proposition}
\newtheorem{corollary}[equation]{Corollary}
\theoremstyle{definition}
\newtheorem{definition}[equation]{Definition}
\newtheorem{construction}[equation]{Construction}
\newtheorem{question}[equation]{Question}

\newtheorem{problem}[equation]{Problem}
\newtheorem{example}[equation]{Example}
\newtheorem{exercise}[equation]{Exercise}
\newtheorem*{answer}{Answer}
\newtheorem*{solution}{Solution}
\newtheorem{remark}[equation]{Remark}

\newtheorem{notation}[equation]{Notation}
\newtheorem{noterm}[equation]{Notation and Terminology}

\newcommand\define[1]{\emph{\textbf{#1}}}

\numberwithin{equation}{section}

 \let\t=\tau

\newcommand{\be}{\begin{equation}}
\newcommand{\ee}{\end{equation}}
\def\ba{\begin{align}} 
\def\ea{\end{align}}
\newcommand{\bea}{\begin{eqnarray}}
\newcommand{\eea}{\end{eqnarray}}
\newcommand{\bx}{\begin{example}}
\newcommand{\ex}{\end{example}}
\newcommand{\bex}{\begin{exercise}}
\newcommand{\eex}{\end{exercise}}
\newcommand{\ban}{\begin{answer}}
\newcommand{\ean}{\end{answer}}
\newcommand{\bt}{\begin{theorem}}
\newcommand{\et}{\end{theorem}}
\newcommand{\bc}{\begin{corollary}}
\newcommand{\ec}{\end{corollary}}
\newcommand{\blem}{\begin{lemma}}
\newcommand{\elem}{\end{lemma}}
\newcommand{\bp}{\begin{problem}}
\newcommand{\ep}{\end{problem}}
\newcommand{\bn}{\begin{proposition}}
\newcommand{\en}{\end{proposition}}
\newcommand{\bd}{\begin{definition}}
\newcommand{\ed}{\end{definition}}
\newcommand{\bcon}{\begin{construction}}
\newcommand{\econ}{\end{construction}}
\newcommand{\bq}{\begin{question}}
\newcommand{\eq}{\end{question}}
\newcommand{\bprf}{\begin{proof}}
\newcommand{\eprf}{\end{proof}}
\newcommand{\br}{\begin{remark}}
\newcommand{\er}{\end{remark}}
\newcommand{\bs}{\begin{solution}}
\newcommand{\es}{\end{solution}}
\newcommand{\beqs}{\begin{eqnarray}}
\newcommand{\eeqs}{\end{eqnarray}}
\newcommand{\bnt}{\begin{noterm}}
\newcommand{\ent}{\end{noterm}}
\newcommand{\bnot}{\begin{notation}}
\newcommand{\enot}{\end{notation}}

\def\R{{{\mathbb R}}}

\def\N{{{\mathbb N}}}

\def\B{{{\mathbb B}}}

\def\VX{\mathcal{X}}
\def\VY{\mathcal{Y}}
\def\VZ{\mathcal{Z}}

\newcommand{\FinPS}{\mathbf{FinPS}}

\newcommand{\stoch}{\;\xy0;/r.25pc/:(-3,0)*{}="1";(3,0)*{}="2";{\ar@{~>}"1";"2"|(1.06){\hole}};\endxy\!}
\newcounter{sarrow}
\newcommand\xstoch[1]{%
\stepcounter{sarrow}%
\mathrel{\begin{tikzpicture}[baseline= {( $ (current bounding box.south) + (0,-0.1ex) $ )}]
\node[inner sep=.5ex] (\thesarrow) {\;$\scriptstyle #1$\;};
\path[draw,{<[scale=1.5,width=3,length=2]}-,decorate,
  decoration={snake,amplitude=0.3mm,segment length=2.1mm,pre=lineto,pre length=1pt}] 
    (\thesarrow.south east) -- (\thesarrow.south west);
\end{tikzpicture}}%
}

\newcounter{sqarrow}

\newcommand{\ds}{\displaystyle}
\newcommand{\ben}{\renewcommand{\theenumi}{\alph{enumi}} 
\renewcommand{\labelenumi}{(\theenumi)}\begin{enumerate}}
\newcommand{\een}{\end{enumerate}}
\newlength\stateheight
\newlength\minimumstatewidth
\tikzset{width/.initial=\minimummorphismwidth}
\tikzset{colour/.initial=white}

\newif\ifblack\pgfkeys{/tikz/black/.is if=black}
\newif\ifwedge\pgfkeys{/tikz/wedge/.is if=wedge}
\newif\ifvflip\pgfkeys{/tikz/vflip/.is if=vflip}
\newif\ifhflip\pgfkeys{/tikz/hflip/.is if=hflip}
\newif\ifhvflip\pgfkeys{/tikz/hvflip/.is if=hvflip}

\pgfdeclarelayer{foreground}
\pgfdeclarelayer{background}
\pgfsetlayers{background,main,foreground}

\def\thickness{0.4pt}

\makeatletter
\pgfkeys{%
  /tikz/on layer/.code={
    \pgfonlayer{#1}\begingroup
    \aftergroup\endpgfonlayer
    \aftergroup\endgroup
  },
  /tikz/node on layer/.code={
    \gdef\node@@on@layer{%
      \setbox\tikz@tempbox=\hbox\bgroup\pgfonlayer{#1}\unhbox\tikz@tempbox\endpgfonlayer\pgfsetlinewidth{\thickness}\egroup}
    \aftergroup\node@on@layer
  },
  /tikz/end node on layer/.code={
    \endpgfonlayer\endgroup\endgroup
  }
}
\def\node@on@layer{\aftergroup\node@@on@layer}
\makeatother

\makeatletter
\pgfdeclareshape{state}
{
    \savedanchor\centerpoint
    {
        \pgf@x=0pt
        \pgf@y=0pt
    }
    \anchor{center}{\centerpoint}
    \anchorborder{\centerpoint}
    \saveddimen\overallwidth
    {
        \pgf@x=3\wd\pgfnodeparttextbox
        \ifdim\pgf@x<\minimumstatewidth
            \pgf@x=\minimumstatewidth
        \fi
    }
    \savedanchor{\upperrightcorner}
    {
        \pgf@x=.5\wd\pgfnodeparttextbox
        \pgf@y=.5\ht\pgfnodeparttextbox
        \advance\pgf@y by -.5\dp\pgfnodeparttextbox
    }
    \anchor{A}
    {
        \pgf@x=-\overallwidth
        \divide\pgf@x by 4
        \pgf@y=0pt
    }
    \anchor{B}
    {
        \pgf@x=\overallwidth
        \divide\pgf@x by 4
        \pgf@y=0pt
    }
    \anchor{text}
    {
        \upperrightcorner
        \pgf@x=-\pgf@x
        \ifhflip
            \pgf@y=-\pgf@y
            \advance\pgf@y by 0.4\stateheight
        \else
            \pgf@y=-\pgf@y
            \advance\pgf@y by -0.4\stateheight
        \fi
    }
    \backgroundpath
    {
       \begin{pgfonlayer}{foreground}
        \pgfsetstrokecolor{black}
        \pgfsetlinewidth{\thickness}
        \pgfpathmoveto{\pgfpoint{-0.5*\overallwidth}{0}}
        \pgfpathlineto{\pgfpoint{0.5*\overallwidth}{0}}
        \ifhflip
            \pgfpathlineto{\pgfpoint{0}{\stateheight}}
        \else
            \pgfpathlineto{\pgfpoint{0}{-\stateheight}}
        \fi
        \pgfpathclose
        \ifblack
            \pgfsetfillcolor{black!50}
            \pgfusepath{fill,stroke}
        \else
            \pgfusepath{stroke}
        \fi
       \end{pgfonlayer}
    }
}

\pgfdeclareshape{my ground}
{
  \inheritsavedanchors[from=rectangle ee]
  \inheritanchor[from=rectangle ee]{center}
  \inheritanchor[from=rectangle ee]{north}
  \inheritanchor[from=rectangle ee]{south}
  \inheritanchor[from=rectangle ee]{east}
  \inheritanchor[from=rectangle ee]{west}
  \inheritanchor[from=rectangle ee]{north east}
  \inheritanchor[from=rectangle ee]{north west}
  \inheritanchor[from=rectangle ee]{south east}
  \inheritanchor[from=rectangle ee]{south west}
  \inheritanchor[from=rectangle ee]{input}
  \inheritanchor[from=rectangle ee]{output}
  \anchorborder{\csname pgf@anchor@my ground@input\endcsname}

  \backgroundpath{
    \pgf@process{\pgfpointadd{\southwest}{\pgfpoint{\pgfkeysvalueof{/pgf/outer xsep}}{\pgfkeysvalueof{/pgf/outer ysep}}}}
    \pgf@xa=\pgf@x \pgf@ya=\pgf@y
    \pgf@process{\pgfpointadd{\northeast}{\pgfpointscale{-1}{\pgfpoint{\pgfkeysvalueof{/pgf/outer xsep}}{\pgfkeysvalueof{/pgf/outer ysep}}}}}
    \pgf@xb=\pgf@x \pgf@yb=\pgf@y
    \pgfpathmoveto{\pgfqpoint{\pgf@xa}{\pgf@ya}}
    \pgfpathlineto{\pgfqpoint{\pgf@xa}{\pgf@yb}}
    \pgfmathsetlength\pgf@xa{.5\pgf@xa+.5\pgf@xb}
    \pgfmathsetlength\pgf@yc{.16666\pgf@yb-.16666\pgf@ya}
    \advance\pgf@ya by\pgf@yc
    \advance\pgf@yb by-\pgf@yc
    \pgfpathmoveto{\pgfqpoint{\pgf@xa}{\pgf@ya}}
    \pgfpathlineto{\pgfqpoint{\pgf@xa}{\pgf@yb}}
    \advance\pgf@ya by\pgf@yc
    \advance\pgf@yb by-\pgf@yc
    \pgfpathmoveto{\pgfqpoint{\pgf@xb}{\pgf@ya}}
    \pgfpathlineto{\pgfqpoint{\pgf@xb}{\pgf@yb}}
  }
}
\makeatother

\tikzset{inline text/.style =
  {text height=1.2ex,text depth=0.25ex,yshift=0.5mm}}
\tikzset{arrow box/.style =
  {rectangle,inline text,fill=white,draw,
    minimum height=5mm,yshift=-0.5mm,minimum width=5mm}}
\tikzset{bubble/.style =
  {inner sep=0mm,minimum width=3mm,minimum height=3mm,
    draw,shape=circle,fill=white}}
\tikzset{dot/.style =
  {inner sep=0mm,minimum width=1mm,minimum height=1mm,
    draw,shape=circle}}
\tikzset{white dot/.style = {dot,fill=white,text depth=-0.2mm}}
\tikzset{scalar/.style = {diamond,draw,inner sep=1pt}}
\tikzset{square/.style =
  {inner sep=0mm,minimum width=2mm,minimum height=2mm,
    draw,shape=rectangle}}

\tikzset{star/.style = {dot,fill=white,text depth=-0.2mm}}
\tikzset{copier/.style = {dot,fill,text depth=-0.2mm}}
\tikzset{fakecopier/.style = {square,fill,text depth=-0.2mm}}
\tikzset{discarder/.style = {my ground,draw,inner sep=0pt,
    minimum width=4.2pt,minimum height=11.2pt,anchor=input,rotate=90}}

\tikzset{xshiftu/.style = {shift = {(#1, 0)}}}
\tikzset{yshiftu/.style = {shift = {(0, #1)}}}
\tikzset{scriptstyle/.style={font=\everymath\expandafter{\the\everymath\scriptstyle}}}

\begin{document}

\begin{center}{\Large An axiomatic characterization of mutual information}\\
James Fullwood\end{center}

\begin{abstract}
We characterize mutual information as the unique map on ordered pairs of random variables satisfying a set of axioms similar to those of Faddeev's characterization of the Shannon entropy. There is a new axiom in our characterization however which has no analogue for 
Shannon entropy, based on the notion of a \emph{Markov triangle}, which may be thought of as a composition of communication channels for which conditional entropy acts functorially. Our proofs are coordinate-free in the sense that no logarithms appear in our calculations.  
\end{abstract}

\tableofcontents

\section{Introduction}

\indent

Axiomatic characterizations of information measures go back to the seminal work of Shannon \cite{Shannon}, providing conceptual insights into their meaning as well as justification for the analytic formulae involved in their definitions. Various characterizations for Shannon entropy, relative entropy, Renyi and Tsallis entropies, von Neumann and Segal entropies, quantum relative entropy as well as other generalized information measures have appeared in the literature \cite{Aczel}\cite{Faddeev}\cite{Furuichi}\cite{Renyi}\cite{Pe92}\cite{Pe93}\cite{leinster2019short}\cite{Ebanks}, and a review of such enterprise in the classical (i.e., non-quantum) setting appears in the survey of Csisz\'{a}r \cite{Csiszar}. More recently, functorial characterizations of information measures from a categorical viewpoint have appeared in the works of Beaz, Fritz and Leinster \cite{BFL}\cite{BaFr14}, as well as our work with Parzygnat \cite{FP1}, who has proved a functorial characterization of the von Neumann entropy \cite{PaEntropy}. An axiomatic approach to entropy in the theory of biodiversity is the subject of the recent book \cite{LeinsterBook} by Leinster. 

In spite of the breadth of the aforementioned results, the mutual information of a pair of random variables seems to be missing from the story. While an operational characterization of mutual information in the context of algorithmic information theory appears in \cite{Zimand}, to the best of our knowledge an axiomatic characterization in the vein of those surveyed by Csisz\'{a}r in \cite{Csiszar} is absent in the literature. It is then the goal of the present work to introduce mutual information into the axiomatic framework.      

Our main result is Theorem~\ref{MIT971X}, where we prove that the mutual information $\mathbb{I}(X,Y)$ of an ordered pair of random variables is the unique function (up to an arbitrary multiplicative factor) on pairs of random variables satisfying the following axioms: 

\begin{enumerate}
\item\label{A1}
\underline{\bf{Continuity}}: If $(X_n,Y_n)\to (X,Y)$, then $\mathbb{I}\left(X,Y\right)=\lim_{n\to \infty}\mathbb{I}\left(X_n,Y_n\right)$.
\item\label{A2}
\underline{\bf{Strong Additivity}}: Given a random variable $X:\Omega\to \VX$ with probability mass function $p:\VX\to [0,1]$, and a collection of pairs of random variables $(Y^x,Z^x)$ indexed by $\VX$, then
\[
\mathbb{I}\left(\bigoplus_{x\in \VX}p(x)(Y^x,Z^x)\right)=\mathbb{I}(X,X)+\sum_{x\in \VX}p(x)\mathbb{I}(Y^x,Z^x).
\]
\item\label{A3}
\underline{\bf{Symmetry}}: $\mathbb{I}(X,Y)=\mathbb{I}(Y,X)$ for every pair of random variables $(X,Y)$.
\item\label{A4}
\underline{\bf{Invariance Under Pullbacks}}: If $\pi:\Omega'\to \Omega$ is a measure-preserving function, then for every pair of random variables $(X,Y)$ with common domain $\Omega$,
\[
\mathbb{I}(X,Y)=\mathbb{I}(X\circ \pi,Y\circ \pi).
\]
\item\label{A5}
\underline{\bf{Weak Functoriality}}: For every Markov triangle $(X,Y,Z)$, 
\[
\mathbb{I}(X,Z)=\mathbb{I}(X,Y)+\mathbb{I}(Y,Z)-\mathbb{I}(Y,Y).
\]
\item\label{A6}
\underline{\bf{Vacuity}}: If $C$ is a constant random variable, then $\mathbb{I}(X,C)=0$.
\end{enumerate}

The fact that mutual information satisfies axioms \ref{A1}, \ref{A3} and \ref{A6} is well known to anybody familiar with mutual information. As we work at the level of random variables as opposed to simply probability distributions (which we do for wider applicability of our results), axiom \ref{A4} is a reflection of the fact that mutual information only depends on probabilities. For axiom \ref{A2}, we define a convex structure on pairs of random variables in such a way that the strong additivity of Shannon entropy is generalized to our context. Axiom \ref{A5} is defined in terms of the notion of \emph{Markov triangle}, a concept we define based on the notion of a `coalescable' composition of communication channels which was introduced in \cite{FP1}. Intuitively, a Markov triangle may be thought of as a composition of noisy channels over which the associated conditional entropy is additive. Moreover, such axioms are sharp in the sense that if any of the axioms are removed then mutual information may not be characterized. In particular, the joint entropy $H(X,Y)$ satisfies axioms \ref{A1}-\ref{A5}, while the conditional entropy $H(Y|X)$ satisfies all the axioms except the symmetry axiom \ref{A3} (note that since $H(X,X)=0$, axiom \ref{A2} in the case of conditional entropy becomes convex linearity). 

In the spirit of the axiomatic approach, we note that logarithms are absent from all calculations in this paper. 

\vspace{0.5cm}

\noindent
\emph{Acknowledgements}: We thank Arthur J. Parzygnat for many useful discussions. 

\section{Mutual Information}
Let $(\Omega,\Sigma, \mu)$ be a probability space, where $\Omega$ is thought of as the set of all possible outcomes of a data generating process, or experiment, $\Sigma$ is a $\sigma$-algebra of measurable subsets of $\Omega$, and $\mu$ is a probability measure.

\bd
A \define{finite random variable} is a surjective function $X:\Omega\to \VX$ such that $\VX$ is a finite set and $X^{-1}(x)\in \Sigma$ for all $x\in \VX$. In such a case, the set $\VX$ is often referred to as the \define{support}, or \define{alphabet} associated with $X$. The \define{probability mass function} of $X$ is the function $p:\VX\to [0,1]$ given by 
\[
p(x)=\mu\left(X^{-1}(x)\right),
\]  
and the \define{Shannon entropy} of $X$ is the non-negative real number $H(X)$ given by
\[
H(X)=-\sum_{x\in \VX}p(x)\log\left(p(x)\right).
\]
The collection of all finite random variables on $\Omega$ will be denoted $\bold{FRV}(\Omega)$.
\ed

\bd
Let $\left(X,Y\right)\in \bold{FRV}(\Omega)\times \bold{FRV}(\Omega)$ be an ordered pair of random variables with supports $\VX$ and $\VY$ respectively. 
\begin{itemize}
 \item
The \define{joint distribution function} of $(X,Y)$ is the function $\vartheta:\VX\times \VY\to [0,1]$ given by
\[
\vartheta(x,y)=\mu\left(X^{-1}(x)\cap Y^{-1}(y)\right),
\]
\item
The \define{joint entropy} of $(X,Y)$ is the non-negative real number given by
\[
H(X,Y)=-\sum_{x\in \VX}\sum_{y\in \VY}\vartheta(x,y)\log\left(\vartheta(x,y)\right),
\]
\item
The \define{mutual information} of $(X,Y)$ is the real number $\mathbb{I}(X,Y)$ given by
\[
\mathbb{I}(X,Y)=H(X)+H(Y)-H(X,Y).
\]
\end{itemize}
\ed

\br
With every pair of random variables $(X,Y)$ one may associate a probability transition matrix $p(y|x)$ given by
\[
p(y|x)=\frac{\vartheta(x,y)}{p(x)},
\] 
where $p:\VX\to [0,1]$ is the probability mass function of $X$. As such, one may view $(X,Y)$ as a noisy channel $\VX\xstoch{} \VY$ together with the prior distribution $p$ on its set of inputs. 
\er

We now list some well-known properties of mutual information which will be useful for our purposes (see e.g. \cite{CoTh06} for proofs). 

\bn\label{MIPROP}
Mutual information satisfies the following properties.
\begin{enumerate}[i.]
\item
$\mathbb{I}(X,Y)\geq 0$ for all $(X,Y)\in \bold{FRV}(\Omega)\times \bold{FRV}(\Omega)$.
\item\label{sym}
$\mathbb{I}(X,Y)=\mathbb{I}(Y,X)$ for all $(X,Y)\in \bold{FRV}(\Omega)\times \bold{FRV}(\Omega)$.
\item
$\mathbb{I}(X,X)=H(X)$ for all $X\in \bold{FRV}(\Omega)$.
\item\label{vanish}
$\mathbb{I}(X,C)=0$ for every constant random variable $C\in \bold{FRV}(\Omega)$.
\end{enumerate}
\en

\bd\label{CANON19}
The \define{canonical product} on $\bold{FRV}(\Omega)$ is the map $\mathscr{P}:\bold{FRV}(\Omega)\times \bold{FRV}(\Omega)\to \bold{FRV}(\Omega)$, given by $\mathscr{P}(X,Y)(\omega)=(X(\omega),Y(\omega))\in \VX\times \VY$ for all $\omega\in \Omega$. 
\ed

\bn\label{PSX19}
Let $(X,Y)\in \bold{FRV}(\Omega)\times \bold{FRV}(\Omega)$. Then the following statements hold.
\begin{enumerate}[i.]
\item \label{ISX1}
The probability mass function of $\mathscr{P}(X,Y)$ is the joint distribution function $\vartheta(x,y)$. In particular, $H(X,Y)=H\left(\mathscr{P}(X,Y)\right)$.
\item
$\mathbb{I}\left(X,\mathscr{P}(X,Y)\right)=H(X)$.
\end{enumerate}
\en

\bprf
\begin{enumerate}[i.]
\item
Let $\nu:\VX\times \VY\to [0,1]$ denote the probability mass function of $\mathscr{P}(X,Y)$. Then for all $(x,y)\in \VX\times \VY$ we have $\mathscr{P}(X,Y)^{-1}(x,y)=X^{-1}(x)\cap Y^{-1}(y)$, thus
\[
\nu(x,y)=\mu\left(\mathscr{P}(X,Y)^{-1}(x,y)\right)=\mu\left(X^{-1}(x)\cap Y^{-1}(y)\right)=\vartheta(x,y),
\]
as desired.
\item
The statement follows from the fact that $H\left(X,\mathscr{P}(X,Y)\right)=H(X,Y)$.
\end{enumerate}
\eprf

\section{Convexity}
We now generalize the notion of a convex combination of probability distributions to the setting of pairs random variables, which will be used to extend the notion of strong additivity for Shannon entropy to mutual information.

\bd
Let $\VX$ be a finite set, and let $p:\VX\to [0,1]$ be a probability distribution on $\VX$. Then $\bigoplus_{x\in \VX}p(x)\left(\Omega,\Sigma,\mu\right)$ is the probability space associated with the triple $\left(\VX\times \Omega,\VX\times \Sigma,p\times \mu\right)$. Now suppose $Y^x\in \bold{FRV}(\Omega)$ is a collection of random variables indexed by $\VX$, and let $q^x:\VY^x\to [0,1]$ denote the probability mass function of $Y^x$. The \define{$p$-weighted convex sum} $\bigoplus_{x\in \VX}p(x)Y^x\in \bold{FRV}\left(\VX\times \Omega\right)$ is the random variable given by
\be\label{cnvx91}
\left(\bigoplus_{x\in \VX}p(x)Y^x\right)(\tilde{x},\omega)=Y^{\tilde{x}}(\omega).
\ee
It then follows that the probability mass function of $\bigoplus_{x\in \VX}p(x)Y^x$ is a function of the form $r:\coprod_{x\in \VX}\VY^x\to [0,1]$, and using the fact that $\coprod_{x\in \VX}\VY^x$ is canonically isomorphic to the set 
\[
\left\{(x,y)\hspace{1mm}|\hspace{1mm} x\in \VX \hspace{2mm} \text{and} \hspace{2mm} y\in \VY^x\right\},
\]
it follows that $r$ is then given by $r(x,y)=p(x)q^x(y)$.
\ed

A reformulation of the strong additivity property for Shannon entropy in terms of the convex structure just introduced for random variables is given by the following proposition.

\bn
Let $\VX$ be a finite set, let $p:\VX\to [0,1]$ be a probability distribution on $\VX$, and suppose $Y^x\in \bold{FRV}(\Omega)$ is a collection of random variables indexed by $\VX$. Then 
\be\label{eq971}
H\left(\bigoplus_{x\in \VX}p(x)Y^x\right)=H(p)+\sum_{x\in \VX}p(x)H(Y^x),
\ee
where $H(p)$ is the Shannon entropy of the probability distribution $p$.
\en

\bn\label{CPX19}
Let $\VX$ be a finite set, let $p:\VX\to [0,1]$ be a probability distribution on $\VX$, and suppose $(Y^x,Z^x)\in \bold{FRV}(\Omega)\times \bold{FRV}(\Omega)$ is a collection of pairs of random variables indexed by $\VX$. Then
\be\label{eqX19}
\bigoplus_{x\in \VX}p(x)\mathscr{P}(Y^x,Z^x)=\mathscr{P}\left(\bigoplus_{x\in \VX}p(x)Y^x,\bigoplus_{x\in \VX}p(x)Z^x\right)
\ee
\en

\bprf
Let $(\tilde{x},\omega)\in \VX\times \Omega$. Then
\begin{eqnarray*}
\left(\bigoplus_{x\in \VX}p(x)\mathscr{P}(Y^x,Z^x)\right)(\tilde{x},\omega)\overset{\eqref{cnvx91}}=\mathscr{P}(Y^{\tilde{x}},Z^{\tilde{x}})(\omega)&=&\left(Y^{\tilde{x}}(\omega),Z^{\tilde{x}}(\omega)\right) \\
&\overset{\eqref{cnvx91}}=&\left(\bigoplus_{x\in \VX}p(x)Y^x(\tilde{x},\omega),\bigoplus_{x\in \VX}p(x)Z^x(\tilde{x},\omega)\right) \\
&=&\mathscr{P}\left(\bigoplus_{x\in \VX}p(x)Y^x,\bigoplus_{x\in \VX}p(x)Z^x\right)(\tilde{x},\omega),
\end{eqnarray*}
thus equation \eqref{eqX19} holds.
\eprf

In light of Proposition~\ref{CPX19}, we make the following definition.

\bd
Let $\VX$ be a finite set, let $p:\VX\to [0,1]$ be a probability distribution on $\VX$, and suppose $(Y^x,Z^x) \in \bold{FRV}(\Omega)\times \bold{FRV}(\Omega)$ is a collection of pairs of random variables indexed by $\VX$. The \define{$p$-weighted convex sum} $\bigoplus_{x\in \VX}p(x)(Y^x,Z^x)\in \bold{FRV}\left(\VX\times \Omega\right)\times \bold{FRV}\left(\VX\times \Omega\right)$ is defined to be the the ordered pair $\left(\bigoplus_{x\in \VX}p(x)Y^x,\bigoplus_{x\in \VX}p(x)Z^x\right)$.
\ed

\bn[Strong Additivity of Mutual Information]
\label{stadd89}
Let $\VX$ be a finite set, let $p:\VX\to [0,1]$ be a probability distribution on $\VX$, and suppose $(Y^x,Z^x)\in \bold{FRV}(\Omega)\times \bold{FRV}(\Omega)$ is a collection of pairs of random variables indexed by $\VX$. Then
\[
\mathbb{I}\left(\bigoplus_{x\in \VX}p(x)(Y^x,Z^x)\right)=H(p)+\sum_{x\in \VX}p(x)\mathbb{I}(Y^x,Z^x),
\] 
where $H(p)$ is the Shannon entropy of the probability distribution $p$.
\en

\bprf
Indeed,
\begin{eqnarray*}
\mathbb{I}\left(\bigoplus_{x\in \VX}p(x)(Y^x,Z^x)\right)&=&\mathbb{I}\left(\bigoplus_{x\in \VX}p(x)Y^x,\bigoplus_{x\in \VX}p(x)Z^x\right) \\
&=&H\left(\bigoplus_{x\in \VX}p(x)Y^x\right)+H\left(\bigoplus_{x\in \VX}p(x)Z^x\right)-H\left(\bigoplus_{x\in \VX}p(x)Y^x,\bigoplus_{x\in X}p(x)Z^x\right) \\
&\overset{\eqref{eqX19}}=&H\left(\bigoplus_{x\in \VX}p(x)Y^x\right)+H\left(\bigoplus_{x\in \VX}p(x)Z^x\right)-H\left(\bigoplus_{x\in \VX}p(x)\mathscr{P}(Y^x,Z^x)\right) \\
&\overset{\eqref{eq971}}=&2H(p)+\sum_{x\in \VX}p(x)\left(H(Y^x)+H(Z^x)\right)-\left(H(p)+\sum_{x\in \VX}p(x)H\left(Y^x,Z^x\right)\right) \\
&=&H(p)+\sum_{x\in \VX}p(x)\left(H(Y^x)+H(Z^x)-H(Y^x,Z^x)\right) \\
&=&H(p)+\sum_{x\in \VX}p(x)\mathbb{I}(Y^x,Z^x),
\end{eqnarray*}
as desired.
\eprf

\section{Continuity}

\bd
Let $X_n \in\bold{FRV}(\Omega)$ be a sequence of random variables, and let $p_n:\VX_n\to [0,1]$ be the associated sequence of probability mass functions. Then $X_n$ is said to \define{weakly converge} (or \define{converge in distribution}) to the random variable $X \in\bold{FRV}(\Omega)$ with probability mass function $p:\VX\to [0,1]$ if the following conditions hold.
\begin{enumerate}[i.]
\item
There exists an $N\in\N$ for which $\VX_{n}=\VX$ for all $n\ge N$.
\item
For all $x\in \VX$ we have $\ds \lim_{n\rightarrow\infty}p_{n}(x)=p(x)$, i.e., $p_n\to p$ pointwise.
\end{enumerate} 
In such a case, we write $X_n\to X$. If $(X_n,Y_n) \in\bold{FRV}(\Omega)\times \bold{FRV}(\Omega)$ is a sequence of pairs of random variables, then $(X_n,Y_n)$ is said to \define{weakly converge} to $(X,Y) \in\bold{FRV}(\Omega)\times \bold{FRV}(\Omega)$ if $\mathscr{P}(X_n,Y_n)\to \mathscr{P}(X,Y)$.
\ed

\bn
Shannon entropy is continuous, i.e., if $X_n\to X$, then 
\[
H\left(X\right)=\lim_{n\to \infty}H(X_n).
\]
\en

\bprf
This result is standard, see e.g. \cite{Faddeev} or \cite{BFL}.
\eprf

\bn\label{MICONT}
Mutual information is continuous, i.e., if $\left(X_n,Y_n\right)\to (X,Y)$,  then  
\[
\mathbb{I}\left(X,Y\right)=\lim_{n\to \infty}\mathbb{I}(X_n,Y_n).
\]
\en

\bprf
Suppose $(X_n,Y_n)\to (X,Y)$, so that $X_n\to X$, $Y_n\to Y$ and $H(X_n,Y_n)\to H(X,Y)$. We then have
\[
\mathbb{I}(X,Y)=H(X)+H(Y)-H(X,Y)=\lim_{n\to \infty}\left(H(X_n)+H(Y_n)-H(X_n,Y_n)\right) =\lim_{n\to \infty}\mathbb{I}(X_n,Y_n), 
\]
as desired.
\eprf

\section{Markov Triangles}
In this section we define the notion of a \emph{Markov triangle}, a concept based on the notion of a `coalescable' composition of communication channels which was introduced in \cite{FP1}. Such a notion will be crucial for our characterization of mutual information.

\bd
Let $X \in\bold{FRV}(\Omega)$ be a random variable with probability mass function $p:\VX\to [0,1]$, and let $x\in \VX$. Then for any random variable $Y \in\bold{FRV}(\Omega)$, the  \define{conditional distribution function} of $Y$ given $X=x$ is the function $q^x:\VY\to [0,1]$ given by
\[
q^x(y)=\begin{cases}
\frac{\vartheta(x,y)}{p(x)} \quad \text{if} \quad p(x)\neq 0 \\
0 \quad \hspace{.72cm} \text{otherwise}.  \\
\end{cases}
\]
From here on, the value $q^x(y)$ will be denoted $q(y|x)$. The \define{conditional entropy} of $Y$ given $X$ is the non-negative real number $H(Y|X)$ given by
\[
H(Y|X)=\sum_{x\in X}p(x)H(q^x),
\]
where $H(q^x)$ is the Shannon entropy of the distribution $q^x$ on $Y$.
\ed

\bn
Let $(X,Y)$ be a pair of random variables. Then 
\be\label{MISX93}
\mathbb{I}(X,Y)=\mathbb{I}(Y,Y)-H(Y|X).
\ee
\en

\bprf
Since $\mathbb{I}(Y,Y)=H(Y)$, the statement follows from the well-known fact that $\mathbb{I}(X,Y)=H(Y)-H(Y|X)$, the proof of which may be found in any information theory text (e.g. \cite{CoTh06}).
\eprf

\bd
Let $(X,Y,Z)$ be a triple of random variables with supports $\VX$, $\VY$ and $\VZ$ respectively, and let  $q(y|x)$, $p(z|y)$ and $r(z|x)$ denote the associated conditional distribution functions. Then $(X,Y,Z)$ is said to form a \define{Markov triangle} if there exists a function $h:\VZ\times \VX\to \VY$ such that for all $(z,x)\in \VZ\times \VX$ we have
\[
r(z|x)=p\left(z|h(z,x)\right)q\left(h(z,x)|x\right).
\]
In such a case, $h$ is said to be a \define{mediator function} for the triple $(X,Y,Z)$. 
\ed

\br
A Markov triangle $(X,Y,Z)$ with supports $\VX$, $\VY$ and $\VZ$ may be thought of as a composition of noisy channels $\VX\xstoch{f}\VY\xstoch{g} \VZ$ such that if $z\in \VZ$ is the output of the channel $g\circ f$, and one is given the information that the associated input was $x\in \VX$, then the output at the intermediary stage $\VY$ was necessarily $y=h(z,x)$ (where $h$ is the associated mediator function). As compositions of deterministic channels always satisfy this property, Markov triangles are a generalization of compositions of deterministic channels. While Markov triangles play a crucial role in our characterization of mutual information and also the characterizations of conditional entropy and information loss in \cite{FP1}, their broader significance in the study of information measures has yet to be determined.  
\er

\bn\label{INCNI}
Suppose $(X,Y,Z)$ is a Markov triangle. Then
\[
\mathbb{I}(X,Z)=\mathbb{I}(X,Y)+\mathbb{I}(Y,Z)-\mathbb{I}(Y,Y).
\]
In particular, $\mathbb{I}(X,Z)\leq \mathbb{I}(X,Y)+\mathbb{I}(Y,Z)$.
\en

Before giving a proof of Proposition~\ref{INCNI}, we first need the following lemma.

\blem\label{MT97X}
Suppose $(X,Y,Z)$ is a Markov triangle. Then
\be\label{FCE89}
H(Z|X)=H(Z|Y)+H(Y|X).
\ee
\elem

\bprf
The statement is simply a reformulation of Theorem~2 in \cite{FP1}.
\eprf

\bprf[Proof of Proposition~\ref{INCNI}]
Suppose $(X,Y,Z)$ is a Markov triangle. Then
\begin{eqnarray*}
\mathbb{I}(X,Z)\overset{\eqref{MISX93}}=\mathbb{I}(Z,Z)-H(Z|X)&\overset{\eqref{FCE89}}=&\mathbb{I}(Z,Z)-\left(H(Z|Y)+H(Y|X)\right) \\
&=&\mathbb{I}(Y,Y)-H(Y|X)+\mathbb{I}(Z,Z)-H(Z|Y)-\mathbb{I}(Y,Y) \\
&\overset{\eqref{MISX93}}=&\mathbb{I}(X,Y)+\mathbb{I}(Y,Z)-\mathbb{I}(Y,Y),
\end{eqnarray*}
as desired.
\eprf

\bn\label{MKVT771}
Let $X,Y\in \bold{FRV}(\Omega)$ be random variables with probability mass functions $p:\VX\to [0,1]$ and $q:\VY\to [0,1]$ respectively. Then the following statements hold.
\begin{enumerate}[i.] 
\item\label{MTX1}
The triple $\left(X,\mathscr{P}(X,Y),Y\right)$ is a Markov triangle.
\item\label{MTX2}
If $f:\VX\to \VX'$ is a bijection, then the triple $(X,f\circ X,Y)$ is a Markov triangle.
\item\label{MTX3}
If $g:\VY\to \VY'$ is a bijection, then the triple $(X,Y,g\circ Y)$ is a Markov triangle.
\end{enumerate}
\en

\bprf
\begin{enumerate}[i.]
\item
Let $r(y|x)$ be the conditional distribution associated with $(X,Y)$, let $p\left(y|(\tilde{x},\tilde{y})\right)$ be the conditional distribution associated with $\left(\mathscr{P}(X,Y),Y\right)$, and let $q\left((\tilde{x},\tilde{y})|x\right)$ be the conditional distribution function associated with $\left(X,\mathscr{P}(X,Y)\right)$. Then for all $y\in \VY$ and $x\in \VX$ we have
\[
r(y|x)=\sum_{(\tilde{x},\tilde{y})\in \VX\times \VY}p\left(y|(\tilde{x},\tilde{y})\right)q\left((\tilde{x},\tilde{y})|x\right)=p\left(y|(x,y)\right)q\left((x,y)|x\right),
\]
where the second equality comes from the fact that $p\left(y|(\tilde{x},\tilde{y})\right)=0$ unless $\tilde{y}=y$ and $q\left((\tilde{x},\tilde{y})|x\right)=0$ unless $x=\tilde{x}$. It then follows that the function $h:\VY\times \VX\to \VX\times \VY$ given by $h(y,x)=(x,y)$ is a mediator function for $\left(X,\mathscr{P}(X,Y),Y\right)$, thus $\left(X,\mathscr{P}(X,Y),Y\right)$ is a Markov triangle.
\item
Let $r(y|x)$ be the conditional distribution associated with $(X,Y)$, let $p\left(y|x'\right)$ be the conditional distribution associated with $\left(f\circ X,Y\right)$, and let $q\left(x'|x\right)$ be the conditional distribution function associated with $\left(X,f\circ X\right)$. Then for all $y\in \VY$ and $x\in \VX$ we have
\[
r(y|x)=\sum_{x'\in \VX'}p(y|x')q(x'|x)=p\left(y|f(x)\right)q\left(f(x)|x\right),
\]
where the second equality comes from the fact that $q(x'|x)=0$ unless $x'=f(x)$. It then follows that the function $h:\VY\times \VX\to \VX'$ given by $h(y,x)=f(x)$ is a mediator function for $(X,f\circ X,Y)$, thus $(X,f\circ X,Y)$ is a Markov triangle.
\item
Let $r(y'|x)$ be the conditional distribution associated with $(X,g\circ Y)$, let $p\left(y'|y\right)$ be the conditional distribution associated with $\left(Y,g\circ Y\right)$, and let $q\left(y|x\right)$ be the conditional distribution associated with $\left(X,Y\right)$. Then for all $y'\in \VY'$ and $x\in \VX$ we have
\[
r(y'|x)=\sum_{y\in \VY}p(y'|y)q(y|x)=p\left(y'|g^{-1}(y')\right)q\left(g^{-1}(y')|x\right),
\]
where the second equality comes from the fact that $p(y'|y)=0$ unless $y=g^{-1}(y')$. It then follows that the function $h:\VY'\times \VX\to \VX'$ given by $h(y',x)=g^{-1}(y')$ is a mediator function for $(X,Y,g\circ Y)$, thus $(X,Y,g\circ Y)$ is a Markov triangle.
\end{enumerate}
\eprf

\section{Characterization Theorem}

We now state and prove our characterization theorem for mutual information.

\bd
Let $(\Omega,\Sigma, \mu)$ and $(\Omega',\Sigma', \mu')$ be probability spaces. A map  $\pi:\Omega'\to \Omega$ is said to be \define{measure-preserving} if for all $\sigma\in \Sigma$ we have $\pi^{-1}(\sigma)\in \Sigma'$ and
\[
\mu'\left(\pi^{-1}(\sigma)\right)=\mu(\sigma).
\]
\ed

\bd
Let $F$ be a map that sends pairs of random variables to the real numbers.
\begin{itemize}
\item
$F$ is said to be \define{continuous} if 
\be\label{eqcont}
F\left(X,Y\right)=\lim_{n\to \infty}F\left(X_n,Y_n\right)
\ee
whenever $(X_n,Y_n)\to (X,Y)$.
\item 
$F$ is said to be \define{strongly additive} if given a random variable $X$ with probability mass function $p:\VX\to [0,1]$, and a collection of pairs of random variables $(Y^x,Z^x)$ indexed by $\VX$, then
\be\label{eqstadd}
F\left(\bigoplus_{x\in \VX}p(x)(Y^x,Z^x)\right)=F(X,X)+\sum_{x\in \VX}p(x)F(Y^x,Z^x).
\ee
\item
$F$ is said to be \define{symmetric} if $F(X,Y)=F(Y,X)$ for every pair of random variables $(X,Y)$.
\item
$F$ is said to be \define{invariant under pullbacks} if for every pair of random variables $(X,Y)\in \bold{FRV}(\Omega)\times \bold{FRV}(\Omega)$ and every measure-preserving map $\pi:\Omega'\to \Omega$ we have 
\be\label{eqinvpb}
F(X,Y)=F(X\circ \pi,Y\circ \pi).
\ee
\item
$F$ is said to be \define{weakly functorial} if for every Markov triangle $(X,Y,Z)$ we have
\be\label{eqmkt}
F(X,Z)=F(X,Y)+F(Y,Z)-F(Y,Y).
\ee
\end{itemize}
\ed

\br
The terminology ``weakly functorial'' comes from viewing \eqref{eqmkt} from a category-theoretic perspective. In particular, with a pair of random variables $(X,Y)$ one may associate a noisy cannel $\VX\xstoch{f} \VY$ where $\VX=\text{Supp}(X)$ and $\VY=\text{Supp}(Y)$, so that a Markov triangle $(X,Y,Z)$ then corresponds to a composition $\VX\xstoch{f} \VY\xstoch{g} \VZ$ with $\VZ=\text{Supp}(Z)$. If $\FinPS$ denotes the category of noisy channels and $\B\R$ denotes the category with one object whose morphisms are the real numbers (with composition corresponding to addition), then a map $F:\FinPS\to \B\R$ is a functor if
\be\label{FEKT197}
F(g\circ f)=F(g)+F(f).
\ee
Rewriting \eqref{FEKT197} in terms of the pairs of random variables for which the morphisms $f$, $g$ and $g\circ f$ are associated with, then the functoriality condition \eqref{FEKT197} reads
\[
F(X,Z)=F(X,Y)+F(Y,Z),
\]
thus the condition $F(X,Z)\leq F(X,Y)+F(Y,Z)$ is a weaker form of functoriality. For more on information measures from a category-theoretic perspective see \cite{BaFr14}\cite{BFL}\cite{FP1}\cite{PaEntropy}.
\er

\bt[Axiomatic Characterization of Mutual Information]\label{MIT971X}
Let $F$ be a map that sends pairs of random variables to the non-negative real numbers, and suppose $F$ satisfies the following conditions.
\begin{enumerate}
\item\label{C1}
$F$ is continuous.
\item\label{C2}
$F$ is strongly additive.
\item\label{C4}
$F$ is symmetric.
\item\label{C3}
$F$ is weakly functorial.
\item\label{C6}
$F$ is invariant under pullbacks.
\item\label{C5}
$F(X,C)=0$ for every constant random variable $C$.
\end{enumerate} 
Then $F$ is a non-negative multiple of mutual information. Conversely, mutual information satisfies conditions \ref{C1}-\ref{C5}.
\et

Before giving a proof we first need several lemmas. The first lemma states that a map $F$ on pairs of random variables which is continuous and invariant under pullbacks only depends on the underlying probability mass functions of the random variables.

\blem\label{LSX735}
Let $F$ be a map from pairs of random variables to the real numbers which is continuous and invariant under pullbacks, and suppose $(X,Y)\in \bold{FRV}(\Omega)\times \bold{FRV}(\Omega)$ and $(X',Y')\in \bold{FRV}(\Omega')\times \bold{FRV}(\Omega')$ are such that the associated joint distribution functions $\vartheta:\VX\times \VY\to [0,1]$ and $\vartheta':\VX\times \VY\to [0,1]$ are equal. Then $F(X,Y)=F(X',Y')$.
\elem

\bprf
Let $\pi:\Omega\times \Omega'\to \Omega$ and $\pi':\Omega\times \Omega'\to \Omega'$ be the natural projections. Since both the natural projections are measure-preserving, we have $F\left(X,Y\right)=F\left(X\circ \pi,Y\circ \pi\right)$, $F\left(X',Y'\right)=F\left(X'\circ \pi',Y'\circ \pi'\right)$, and moreover, from the assumption that $\vartheta=\vartheta'$ it follows that the joint distribution functions associated with $\left(X\circ \pi,Y\circ \pi\right)$ and $\left(X'\circ \pi',Y'\circ \pi'\right)$ are equal. It then follows that if $(X_n,Y_n)$ is the constant sequence given by $X_n=X'\circ \pi'$ and $Y_n=Y'\circ \pi'$ for all $n\in \N$, then $(X_n,Y_n)\to (X\circ \pi,Y\circ \pi)$ (since $\mathscr{P}(X_n,Y_n)\to \mathscr{P}(X\circ \pi,Y\circ \pi)$). We then have
\[
F(X,Y)\overset{\eqref{eqinvpb}}=F\left(X\circ \pi,Y\circ \pi\right)\overset{\eqref{eqcont}}=\lim_{n\to \infty}F(X_n,Y_n)=F\left(X'\circ \pi',Y'\circ \pi'\right)\overset{\eqref{eqinvpb}}=F\left(X',Y'\right),
\]
as desired.
\eprf

\blem\label{FUCT971}
Let $X$ be a random variable with probability mass function $p:\VX\to [0,1]$, let $f:\VX\to \VY$ be a bijection, and suppose $C$ is a constant random variable. Then the following statements hold.
\begin{enumerate}[i.]
\item\label{K1}
The triples $\left(X,f\circ X,C\right)$ and $\left(f\circ X,X,C\right)$ are both Markov triangles.
\item\label{K2}
Let $F$ be a map which sends pairs of random variables to real numbers, and suppose $F$ is symmetric and weakly functorial. Then 
\end{enumerate} 
\be\label{esX747}
F(X,C)-F\left(f\circ X,C\right)+F\left(f\circ X,f\circ X\right)=F\left(f\circ X,C\right)-F(X,C)+F(X,X).
\ee
\elem

\bprf
\begin{enumerate}[i.]
\item The statement follows from item \ref{MTX2} of Proposition~\ref{MKVT771}.
\item
By item \ref{K1}, the triples $\left(X,f\circ X,C\right)$ and $\left(f\circ X,X,C\right)$ are both Markov triangles, thus the weak functoriality of $F$ yields 
\be\label{esX748}
F(X,C)=F(X,f\circ X)+F(f\circ X,C)-F(f\circ X,f\circ X),
\ee
and
\be\label{esX749}
F(f\circ X,C)=F(f\circ X,X)+F(X,C)-F(X,X).
\ee
And since $F$ is symmetric $F(X,f\circ X)=F(f\circ X,X)$, thus equations \eqref{esX748} and \eqref{esX749} imply equation \eqref{esX747}, as desired.
\end{enumerate}
\eprf

The next lemma is Baez, Fritz and Leinster's reformulation of Faddeev's characterization of Shannon entropy \cite{Faddeev}, which they use in their characterization of the information loss associated with a deterministic mapping \cite{BFL}. This lemma will allow us to relate $F(X,X)$ to the Shannon entropy $H(X)$.

\blem \label{FTX77}
Let $\mathcal{S}$ be a map which sends finite probability distributions to the non-negative real numbers, and suppose $\mathcal{S}$ satisfies the following conditions.
\begin{enumerate}[i.]
\item\label{FT1}
$\mathcal{S}$ is continuous, i.e., if $p_n:\VX\to [0,1]$ is a convergent sequence of probability distributions on a finite set $\VX$ (i.e., if $\lim_{n\to \infty}p_n(x)$ exists for all $x\in \VX$), then 
\[
\mathcal{S}\left(\ds \lim_{n\to \infty}p_n\right)=\ds \lim_{n\to \infty}\mathcal{S}(p_n).
\]
\item\label{FT2}
$\mathcal{S}(1)=0$ for the distribution $1:\{\star\}\to [0,1]$.
\item\label{FT3}
If $q:\VY\to [0,1]$ is a probability distribution on a finite set $\VY$ and $f:\VX\to \VY$ is a bijection, then $\mathcal{S}(q)=\mathcal{S}(q\circ f)$.
\item\label{FT4}
If $p:\VX\to [0,1]$ is a probability distribution on a finite set $\VX$, and $q^x:\VY^x\to [0,1]$ is a collection of finite probability distributions indexed by $\VX$, then
\[
\mathcal{S}\left(\bigoplus_{x\in \VX}p(x)q^x\right)=\mathcal{S}(p)+\sum_{x\in \VX}p(x)\mathcal{S}(q^x),
\]
where $\bigoplus_{x\in \VX}p(x)q^x:\coprod_{x\in \VX}\VY^x\to [0,1]$ is the finite distribution given by $\left(\bigoplus_{x\in \VX}p(x)q^x\right)(\tilde{x},y_{\tilde{x}})=p(\tilde{x})q^{\tilde{x}}(y_{\tilde{x}})$.
\end{enumerate}
Then $\mathcal{S}$ is a non-negative multiple of Shannon entropy.
\elem

\blem\label{AC91SX}
Let $F$ be a map which sends pairs of random variables to the non-negative real numbers satisfying conditions \ref{C1}-\ref{C5} of Theorem~\ref{MIT971X}, and let $\mathscr{E}$ be the map on random variables given by 
\[
\mathscr{E}(X)=F(X,X).
\]
Then $\mathscr{E}$ is a non-negative multiple of Shannon entropy. 
\elem

\bprf
Let $\phi$ be the map which takes a random variable to its probability mass function, let $\sigma$ be a section (so that $\phi\circ \sigma$ is the identity), and let $\mathcal{S}=\mathscr{E}\circ \sigma$. Since $F$ is invariant under pullbacks (condition \ref{C6} of Theorem~\ref{MIT971X}) Lemma~\ref{LSX735} holds, thus the map $\mathcal{S}$ is independent of the choice of a section $\sigma$ of $\phi$, and as such, it follows that $\mathscr{E}=\mathcal{S}\circ \phi$. We now show that $\mathcal{S}$ satisfies items \ref{FT1}-\ref{FT4} of Lemma~\ref{FTX77}, which then implies $\mathscr{E}(X)$ is a non-negative multiple of the Shannon entropy $H(X)$.

\underline{Item \ref{FT1}}: 
Let $p_n:\VX\to [0,1]$ be a sequence of probability distributions on a finite set $\VX$, and suppose $\ds \lim_{n\to \infty}p_n=p$. It then follows that $X_n=\sigma(p_n)$ weakly converges to $X=\sigma(p)$, thus 
\begin{eqnarray*}
\mathcal{S}\left(\ds \lim_{n\to \infty}p_n\right)&=&\mathcal{S}\left(p\right)=(\mathscr{E}\circ \sigma)(p)=\mathscr{E}(X)=F(X,X)=\lim_{n\to \infty}F(X_n,X_n) \\
&=&\lim_{n\to \infty}\mathscr{E}(X_n)=\lim_{n\to \infty}\mathscr{E}(\sigma(p_n))=\lim_{n\to \infty}\mathcal{S}(p_n) \\
\end{eqnarray*}
where the fifth equality follows from the continuity assumption on $F$ (condition \ref{C1} of Theorem~\ref{MIT971X}).

\underline{Item \ref{FT2}}: Let $1:\{\star\}\to [0,1]$ be a point mass distribution, so that $\sigma(1)=C$ with $C$ a constant random variable. Then $\mathcal{S}(1)=\mathscr{E}(\sigma(1))=\mathscr{E}(C)=F(C,C)=0$, where the last equality follows from condition \ref{C5} of Theorem~\ref{MIT971X}, i.e., that $F(X,C)=0$ for every constant random variable $C$.

\underline{Item \ref{FT3}}: Let $X$ be a random variable with probability mass function $p:\VX\to [0,1]$, and suppose $f:\VX\to \VY$ is a bijection. Since $F$ is symmetric and weakly functorial (conditions \ref{C4} and \ref{C3} of Theorem~\ref{MIT971X}), the hypotheses of item \ref{K2} Lemma~\ref{FUCT971} are satisfied, so that equation \eqref{esX747} holds, i.e., for any constant random variable $C$ we have
\[
F(X,C)-F\left(f\circ X,C\right)+F\left(f\circ X,f\circ X\right)=F\left(f\circ X,C\right)-F(X,C)+F(X,X).
\]
And since $F(X,C)=F\left(f\circ X,C\right)=0$ by condition \ref{C5} of Theorem~\ref{MIT971X}, it follows that $F(X,X)=F(f\circ X,f\circ X)$. Now let $q:\VY\to [0,1]$ be the probability mass function of $f\circ X$, so that $q=p\circ f^{-1}$. We then have
\[
\mathcal{S}(p)=\mathscr{E}(X)=F(X,X)=F(f\circ X,f\circ X)=\mathscr{E}(f\circ X)=\mathcal{S}(q)=\mathcal{S}(p\circ f^{-1}),
\]
thus $\mathcal{S}$ satisfies item \ref{FT3} of Faddeev's Theorem.

\underline{Item \ref{FT4}}: Let $X$ be a random variable with probability mass function $p:\VX\to [0,1]$, $Y^x$ a collection of random variables indexed by $\VX$, and let $q^x:\VY^x\to [0,1]$ be the associated probability mass functions for all $x\in \VX$. Then $\bigoplus_{x\in \VX}p(x)Y^x$ has probability mass function $\bigoplus_{x\in \VX}p(x)q^x$, thus
\begin{eqnarray*}
\mathcal{S}\left(\bigoplus_{x\in \VX}p(x)q^x\right)=\mathscr{E}\left(\bigoplus_{x\in \VX}p(x)Y^x\right)&=&F\left(\bigoplus_{x\in \VX}p(x)Y^x,\bigoplus_{x\in \VX}p(x)Y^x\right) \\
 &\overset{\eqref{eqX19}}=&F\left(\bigoplus_{x\in \VX}p(x)\left(Y^x,Y^x\right)\right) \\
&=&F(X,X)+\sum_{x\in \VX}p(x)F\left(Y^x,Y^x\right) \\
&=&\mathscr{E}(X)+\sum_{x\in \VX}p(x)\mathscr{E}(Y^x) \\
&=&\mathcal{S}(p)+\sum_{x\in \VX}p(x)\mathcal{S}(q^x),
\end{eqnarray*}
where the fourth equality follows from the strong additivity of $F$, i.e., condition \ref{C2} of Theorem~\ref{MIT971X}. It then follows that $\mathcal{S}$ satisfies item \ref{FT4} of Faddeev's Theorem, as desired.
\eprf

The next lemma is the analogue of property \ref{FT3} of Lemma~\ref{FTX77} for information measures on pairs of random variables.

\blem\label{LXS17}
Let $X,Y\in \bold{FRV}(\Omega)$ be random variables with probability mass functions $p:\VX\to [0,1]$ and $q:\VY\to [0,1]$ respectively, and suppose $F$ is a map on pairs of random variables to the real numbers which is symmetric, weakly functorial, and $F(X,C)=0$ for every constant random variable $C$. If $f:\VX\to \VX'$ and $g:\VY\to \VY'$ are bijections, then
\be\label{eqphx}
F(X,Y)=F\left(f\circ X,g\circ Y\right).
\ee
\elem

\bprf
Since $f$ is a bijection, $(X,f\circ X,X)$ is a Markov triangle by item \ref{MTX2} of Proposition~\ref{MKVT771}, thus
\be\label{lxs71}
F(X,X)=F(X,f\circ X)+F(f\circ X,X)-F(f\circ X,f\circ X).
\ee
From the proof of Lemma~\ref{AC91SX} it follows that if $F$ is weakly functorial, symmetric and $F(X,C)=0$ for every constant random variable $C$, $F(f\circ X,f\circ X)=F(X,X)$. Moreover by the symmetry of $F$ we have $F(X,f\circ X)=F(f\circ X,X)$, thus equation \eqref{lxs71} implies $F(X,X)=F(X,f\circ X)$.

Now consider the triples $(f\circ X,X,g\circ Y)$ and $(X,Y,g\circ Y)$, which are both Markov triangles by items \ref{MTX2} and \ref{MTX3} of Proposition~\ref{MKVT771}. The weakly functorial assumption on $F$ then yields
\begin{eqnarray*}
F\left(f\circ X,g\circ Y\right)&=&F\left(f\circ X,X\right)+F\left(X,g\circ Y\right)-F\left(X,X\right) \\
&=&F\left(f\circ X,X\right)+\left(F\left(X,Y\right)+F\left(Y,g\circ Y\right)-F\left(Y,Y\right)\right)-F\left(X,X\right), \\
\end{eqnarray*}
and since $F\left(f\circ X,X\right)=F\left(X,X\right)$ and $F\left(Y,g\circ Y\right)=F\left(Y,Y\right)$, it follows that $F(X,Y)=F\left(f\circ X,g\circ Y\right)$, as desired.
\eprf

The next lemma together with the fact that $(X,\mathscr{P}(X,Y),Y)$ is a Markov triangle (by Proposition~\ref{MKVT771}) is the crux of the proof, as we will soon see. 

\blem\label{LSX747}
Let $F$ be a map from pairs of random variables to the real numbers satisfying conditions \ref{C1}-\ref{C5} of Theorem~\ref{MIT971X}, and let $(X,Y)$ be a pair of random variables. Then $F\left(X,\mathscr{P}(X,Y)\right)=F(X,X)$ and $F\left(\mathscr{P}(X,Y),Y\right)=F(Y,Y)$.
\elem

\bprf
Let $p:\VX\to [0,1]$ and $q:\VY\to [0,1]$ be the probability mass functions of $X$ and $Y$ respectively, and for all $x\in \VX$, let $Y^x$ be a random variable with probability mass function $q^x:\VY\to [0,1]$ given by $q^x(y)=q(y|x)$, so that $q^x$ is the conditional distribution of $Y$ given $X=x$. By pulling back to larger sample spaces if necessary, we can assume without loss of generality that each $Y^x\in \bold{FRV}(\Omega)$ for some fixed $\Omega$. We also let $C^x\in \bold{FRV}(\Omega)$ be the constant random variable supported on $\{x\}$ for all $x\in \VX$, we let $f:\coprod_{x\in \VX}\{x\}\to \VX$ and $g:\coprod_{x\in \VX}\VY\to \VX\times \VY$ be the canonical bijections, and we let $\pi:\VX\times \Omega\to \Omega$ be the natural projection. It then follows that $f\circ \bigoplus_{x\in \VX}p(x)C^x$ and $X\circ \pi$ both have probability mass function $p:\VX\to [0,1]$, and also, that $g\circ \bigoplus_{x\in \VX}p(x)Y^x$ and $\mathscr{P}(X,Y)\circ \pi$ both have probability mass function equal to the joint distribution function $\vartheta:\VX\times \VY\to [0,1]$ associated with $(X,Y)$, thus Lemma~\ref{LSX735} yields
\be\label{eqsx997}
F\left(X\circ \pi,\mathscr{P}(X,Y)\circ \pi\right)=F\left(f\circ \bigoplus_{x\in \VX}p(x)C^x,g\circ \bigoplus_{x\in \VX}p(x)Y^x\right).
\ee
We then have
\begin{eqnarray*}
F\left(X,\mathscr{P}(X,Y)\right)\overset{\eqref{eqinvpb}}=F\left(X\circ \pi,\mathscr{P}(X,Y)\circ \pi\right)&\overset{\eqref{eqsx997}}=&F\left(f\circ\bigoplus_{x\in \VX}p(x)C^x,g\circ\bigoplus_{x\in \VX}p(x)Y^x\right) \\
&\overset{\eqref{eqphx}}=&F\left(\bigoplus_{x\in \VX}p(x)C^x,\bigoplus_{x\in \VX}p(x)Y^x\right) \\
&\overset{\eqref{eqX19}}=&F\left(\bigoplus_{x\in \VX}p(x)(C^x,Y^x)\right) \\
&\overset{\eqref{eqstadd}}=&F(X,X)+\sum_{x\in \VX}p(x)F(C^x,Y^x) \\
&=&F(X,X),
\end{eqnarray*}
where the last equality follows from the fact that $F(C,X)=0$ for every constant random variable $C$ (since $F$ is symmetric and $F(X,C)=0$ for every constant random variable $C$). 

As for $F\left(\mathscr{P}(X,Y),Y\right)$, first note that $F\left(Y,\mathscr{P}(Y,X)\right)=F(Y,Y)$ by what what we have just proved. We then have
\[
F\left(\mathscr{P}(X,Y),Y\right)=F\left(Y,\mathscr{P}(X,Y)\right)=F\left(Y,\mathscr{P}(Y,X)\right)=F(Y,Y),
\]
where the first and second equalities follow from symmetry and invariance under pullbacks.
\eprf

\bprf[Proof of Theorem~\ref{MIT971X}]
Suppose $F$ is a map from pairs of random variables to the non-negative real numbers satisfying conditions \ref{C1}-\ref{C5} of Theorem~\ref{MIT971X}. By Lemma~\ref{AC91SX}, there exists a constant $c\geq 0$ such that $F(X,X)=cH(X)$ for all random variables $X$. Now let $(X,Y)$ be an arbitrary pair of random variables. By Proposition~\ref{MKVT771}, the triple $(X,\mathscr{P}(X,Y),Y)$ is a Markov triangle, thus
\begin{eqnarray*}
F(X,Y)&\overset{\eqref{eqmkt}}=&F\left(X,\mathscr{P}(X,Y)\right)+F\left(\mathscr{P}(X,Y),Y\right)-F\left(\mathscr{P}(X,Y),\mathscr{P}(X,Y)\right) \\
&=&F(X,X)+F(Y,Y)-cH\left(\mathscr{P}(X,Y)\right) \\
&=&cH(X)+cH(Y)-cH(X,Y) \\
&=&c\mathbb{I}(X,Y),
\end{eqnarray*}
where the second equality follows from Lemma~\ref{LSX747} and Lemma~\ref{AC91SX}, and the third equality follows from Lemma~\ref{AC91SX} and item \ref{ISX1} of Proposition~\ref{PSX19}, thus $F$ is a non-negative multiple of mutual information. 

Conversely, mutual information satisfies condition \ref{C1} of  by Proposition~\ref{MICONT}, condition \ref{C2} by Proposition~\ref{stadd89}, condition \ref{C4} by item \ref{sym} of Proposition~\ref{MIPROP}, condition \ref{C3} by Proposition~\ref{INCNI}, condition \ref{C6} by the fact that mutual information only depends on probabilities, and condition \ref{C5} by item \ref{vanish} of Proposition~\ref{MIPROP}.
\eprf

\addcontentsline{toc}{section}{\numberline{}Bibliography}
\bibliographystyle{plain}
\bibliography{AXM}

\end{document}